\documentclass[9pt,twocolumn,twoside]{osajnl}

\journal{ao} 

\newcommand{\rtHz}{\sqrt{\text{Hz}}}
\newcommand{\SiN}{Si$_3$N$_4$ }
\renewcommand{\t}[1]{\mathrm{#1}}

\setboolean{shortarticle}{true}

\title{Towards cavity-free ground state cooling of an acoustic-frequency silicon nitride membrane}

\author[1]{Christian Pluchar}
\author[1]{Aman Agrawal}
\author[2]{Edward Schenk}
\author[1,2]{Dalziel J. Wilson}

\affil[1]{College of Optical Sciences, University of Arizona, 1630 E. University Blvd., Tucson AZ, 85721, USA}
\affil[2]{Department of Physics, University of Arizona, 1118 E. Fourth St., Tucson AZ, 85721, USA}

\affil[*]{Corresponding author: dalziel@optics.arizona.edu}

\begin{abstract}
We demonstrate feedback cooling of a millimeter-scale, 40 kHz SiN membrane from room temperature to 5 mK (3000 phonons) using a Michelson interferometer, and discuss the challenges to ground state cooling without an optical cavity.  This advance appears within reach of current membrane technology, positioning it as a compelling alternative to levitated systems for quantum sensing and fundamental weak force measurements.
\end{abstract}

\setboolean{displaycopyright}{true}

\begin{document}

\maketitle


\noindent Strained thin films resonators (strings and membranes) with millimeter dimensions can support acoustic frequency modes with extremely high quality factors, leveraging the effect of dissipation dilution \cite{tsaturyan2017ultracoherent, ghadimi2018elastic,norte2016mechanical, reinhardt2016ultralow}.  It has been speculated that they may enable room temperature quantum optomechanics \cite{norte2016mechanical}, ultra-precise force and acceleration sensing \cite{fischer2019spin, krause_high-resolution_2012}, quantum memories, and detection of fundamental weak signals such as spontaneous waveform collapse \cite{nimmrichter2014optomechanical} and ultralight dark matter \cite{carney2019ultralight}.

Here we discuss an additional potential of acoustic frequency thin film resonators, which is to remove the need for a cavity in quantum optomechanics experiments.  This possibility is due to the large  zero-point fluctuations of an acoustic frequency nanomechanical resonator, as exploited in experiments with levitated nanoparticles \cite{yin2013optomechanics}.  In contrast to levitated nanoparticles, thin film resonators can be read out with high efficiency 
by direct reflection or near-field sensing \cite{barg2017measuring}.  The main challenge to reaching the quantum regime is technical noise, such as laser relaxation-oscillations, which can be far in excess of shot noise at acoustic frequencies.  Optical absorption can also lead to large bolometric effects in a tethered nanostructures, while in principle they can be decoupled from motion of a levitated particle \cite{hebestreit2018measuring}.

To illustrate the potential for "cavity-free" quantum optomechanics, we describe an experiment in which the fundamental mode of a 2.5 mm, high stress silicon nitride (Si$_3$N$_4$) trampoline resonator \cite{norte2016mechanical, reinhardt2016ultralow} is subject to radiation pressure feedback cooling using a Michelson interferometer.  The conditions for ground state cooling are two-fold \cite{wilson2015measurement}: (1) the oscillator's thermal decoherence rate $\Gamma_\t{th}$ must not exceed its frequency $\Omega_0$
\begin{equation}\label{eq:1}
    \Gamma_\t{th} = \frac{k_\t{B}T_0}{\hbar Q_0}<\Omega_0
\end{equation}
and (2) the measurement imprecision $S_{xx}^\t{imp}$ (expressed as a single-sided power spectral density) must be low enough to resolve zero-point motion $x_\t{zp}$ in the thermal decoherence time
\begin{equation}\label{eq:2}
    S_{xx}^\t{imp,gs}= \frac{4 x_\t{zp}^2}{\Gamma_\t{th}} = \frac{2\hbar^2}{k_\t{B}T_0}\frac{Q_0}{m\Omega_0}
\end{equation}
where $T_0$ is the intrinsic device temperature.

 In our experiment, operated at room temperature, an optimized trampoline design \cite{reinhardt2016ultralow,norte2016mechanical} yields a fundamental frequency of $\Omega_0 = 2\pi\cdot 40$ kHz, a quality factor $Q_0 = 3\times 10^7$, and an effective mass of $m = 12$ ng, corresponding to a thermal decoherence rate of $\Gamma_\t{th}= 5\Omega_0$, a zero-point displacement of $x_\t{zp}= 4\;\t{fm}$, and a ground state cooling requirement of (\eqref{eq:2}) $(S_{xx}^\t{imp,gs})^{1/2}\approx 10^{-17}\;\t{m}/\sqrt{\t{Hz}}$. The latter is three orders of magnitude below the sensitivity of our microscope; nevertheless, using an auxiliary laser field as a radiation pressure actuator, we realize feedback cooling to an effective temperature of $T_\t{eff}=5\;\t{mK}$, corresponding to a mean phonon number of
 \begin{equation}\label{eq:3}
 \langle n\rangle=\frac{k_B T_\t{eff}}{\hbar\Omega_0}\approx \sqrt{\frac{S_{xx}^\t{imp}}{S_{xx}^\t{imp,gs}}}\approx 3\times 10^3
 \end{equation}
 Below we discuss the design and limitations of this experiment and speculate that $\langle n\rangle\sim 1$ should be possible with simple modifications, including pre-cooling in a microscopy cryostat and using a common path interferometer topology.  

\section{Measurement-based feedback cooling}

In feedback cooling protocols, a continuous position measurement is used to suppress the thermal motion of a mechanical oscillator by derivative feedback (velocity damping).  The technique dates back to collimation of particle accelerators \cite{van1985stochastic} and is commonly used in atomic force microscopes to improve their dynamic range \cite{poggio_feedback_2007}.  
More recently, feedback cooling has received attention in the cavity optomechanics community as a means to prepare a nanomechanical oscillator in its  ground state. 
Cooling to $\langle n \rangle \sim 4$ has been achieved with with a free space optically levitated nanoparticle \cite{tebbenjohanns2020motional}, while cavity-enhanced measurements have been used to cool a Si$_3$N$_4$ nanostring to $\langle n \rangle\sim4$ \cite{wilson2015measurement} and, more recently, a Si$_3$N$_4$ membrane to $\langle n \rangle\sim 0.3$ \cite{rossi2018measurement}.

An important feature of feedback cooling is that, unlike optomechanical sideband cooling \cite{wilson2007theory, marquardt2007quantum}, it does not require a "good" (sideband-resolved) cavity to reach the ground state \cite{genes_ground-state_2008,Jacobs_comparing_2015}.  It is only necessary to achieve a sufficient measurement  efficiency, which in fact requires a "bad" cavity and in principle requires no cavity at all.  To this end, agnostic to the measurement scheme, consider a real-time estimate $y$ of the oscillator displacement $x$ obscured by imprecision noise $x_\t{imp}$:
\begin{equation}
    y = x+x_\t{imp}.
\end{equation}
Feedback cooling can be understood by including a velocity-proportional feedback force in the Langevin equation
\begin{equation}
    m\ddot{x}+m\Gamma_\t{0}\dot{x}+m\Omega_\t{0}^2x = \sqrt{2k_B T_0m\Gamma_0}\xi(t) - gm\Gamma_0 \dot{y}
\end{equation}
where $\xi(t)$ is a normalized Gaussian white noise process and $\Gamma_0 = \Omega_0/Q_0$ is the intrinsic mechanical damping rate. 
Applying the Wiener-Khinchin theorem, 
the spectral density of physical $(x)$ and apparent $(y)$ displacement can be expressed as \cite{wilson2015measurement,rossi2018measurement}
\begin{subequations}\begin{align}\label{eq:6}
\frac{S_{xx}[\Omega]}{2S_{xx}^\t{zp}}&= |\chi_{g}[\Omega]|^2\left(n_\t{th}+g^2 n_\t{imp}\right)\\
\frac{S_{yy}[\Omega]}{2S_{xx}^\t{zp}}&= |\chi_{g}[\Omega]|^2\left(n_\t{th}+(1+g)^2|\chi_{0}[\Omega]|^{-2} n_\t{imp}\right)
\end{align}\end{subequations}
where $S_{xx}^\t{zp} = 4x_\t{zp}^2/\Gamma_0$ is the zero-point displacement spectral density, $n_\t{th}=k_\t{B}T_0/m\Omega_0$ is the thermal bath occupation, and 
$\chi_g^{-1}\approx (1+g)+2i(\Omega-\Omega_0)\slash\Gamma_0$ 
is the closed-loop mechanical susceptibility. Evidently feedback damping can be ``cold'' in the sense that $n_\t{imp} = S_{xx}^\t{imp}/2S_{xx}^\t{zp}< n_\t{th}$ when the measurement resolves the thermal motion.  Increasing the feedback gain $g$ thus reduces the average displacement of the oscillator $\langle x^2\rangle = \int S_{xx}[\Omega]/2\pi$, resulting in a mean phonon number of 
\begin{align}\label{eq:7}
    \langle n\rangle + \frac{1}{2}= \frac{\langle x^2 \rangle}{2x_{zp}^2}=\frac{n_\t{th}+g^2 n_\t{imp}}{1+g}\ge2\sqrt{n_\t{th}n_\t{imp}}.
\end{align}
Ground state cooling requires accounting for measurement back-action $n_\t{th}\rightarrow n_\t{th}+n_\t{ba}=n_\t{th}+\eta/(16 n_\t{imp})$, where $\eta\in[0,1]$ is the measurement efficiency \cite{wilson2015measurement,rossi2018measurement}.  \eqref{eq:7} thus yields \eqref{eq:2} for $\langle n\rangle < 1$ and \eqref{eq:3} for $1\ll\langle n\rangle\ll n_\t{th}$ (noting that $\Gamma_\t{th} = \Gamma_0 n_\t{th}$).  In addition to high efficiency, we emphasize that reaching low occupancy is facilitated by having a high $Q/m\Omega_0$ factor, which is equivalent to a high force sensitivity $S_{FF}^\t{th} = 4k_{B}Tm\Omega_0/Q$. 

\section{Trampoline Resonator}

 Our mechanical resonator is a modified version of the Si$_3$N$_4$  trampoline introduced by \mbox{Reinhardt} \cite{reinhardt2016ultralow} and Norte \emph{et. al.} \cite{norte2016mechanical}. 
  Trampoline resonators, like strings \cite{ghadimi2017radiation,villanueva2014evidence}, exhibit quality factors scaling as $Q\propto Q_{mat}(h)\sqrt{\sigma}L/h$, where $Q_\t{mat}^{-1}$ is the material loss tangent,  $L$ is the tether length, $h$ is the film thickness, and $\sigma$ is the tensile stress in the film.  Since $\Omega_0\propto \sqrt{\sigma}/L$ and $m\propto hL$, the implication is that $Q/m\Omega_0\propto Q_\t{mat}(h)L/h^2$. Counterintuitively, \emph{larger} devices can have larger zero-point fluctuations.
  
\begin{figure}[t]\centering
\includegraphics[width=\linewidth]{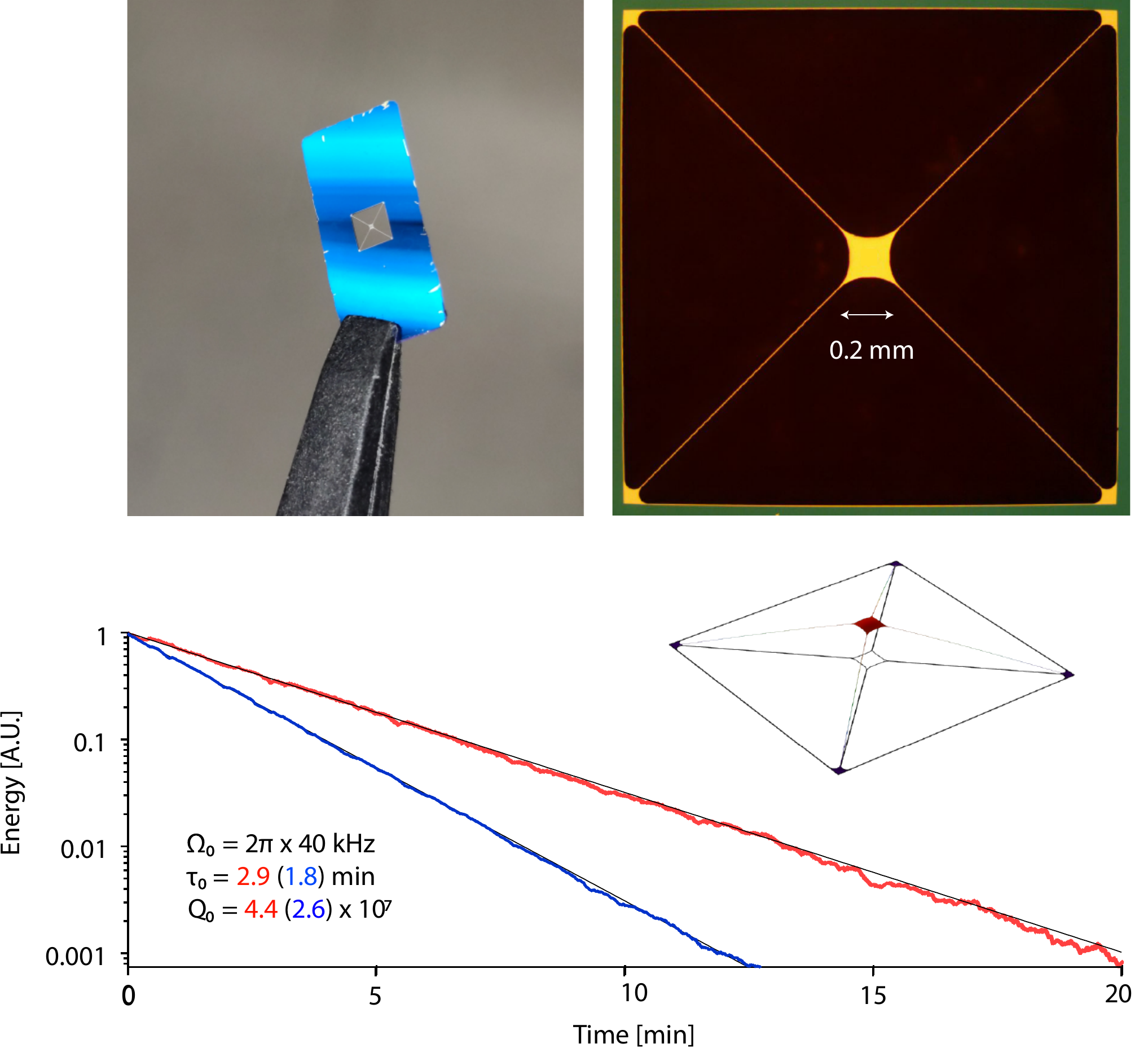}
\caption{\SiN trampoline resonator.  (Top left) Camera image of a typical device.  (Top right) Microscope image of the trampoline used in the experiment.  (Bottom right) Finite element similuation of the fundamental 40 kHz vibrational mode.  (Bottom left) Energy ringdown of the fundamental mode before (red) and after (blue) deposition of a dust particle onto a tether.}
\label{fig:trampoline}
\end{figure}

The trampoline used in our experiment is shown in Fig. \ref{fig:trampoline}.  The device is suspended from a $h \approx 90$ nm thick \SiN film on a 200 $\mu$m thick Si wafer (WaferPro) using a standard two-sided photolithography and wet etching technique \cite{reinhardt2016ultralow,norte2016mechanical,fabnote}. A 200 $\mu$m pad and $L\approx 1.7$ mm long, $w=4.2\,\mu$m wide tethers were chosen, as well as an ``optimal'' \cite{sadeghi2019influence} $50\,\mu$m radius fillet for this window size (2.5 mm) and tether width.  Mechanical ringdown measurements performed in high vacuum ($<10^{-7}$ mbar, Fig. \ref{fig:trampoline}c) reveal a fundamental frequency of $\Omega_0=2\pi\times 39.9$ kHz and a quality factor as high as $Q= 4.4\times 10^7$, which agrees with finite element simulations (COMSOL) assuming a film stress of $\sigma = 0.9$ GPa and internal quality factor of $Q_\t{mat} = 6\times 10^3$.  (The latter is consitent with the $Q_{mat}\propto h$ surface loss model of Villanueva \emph{et. al.} \cite{villanueva2014evidence}, suggesting our device is not limited by clamping loss.)  For the experiments described below, dust deposited on the tether resulted in a reduced quality factor of $Q = 2.6\times 10^7$.  Together with a simulated effective mass of $m=12$ ng, this implies a force sensitivity of $S_{FF}^\t{th}=(43\;\t{aN}/\rtHz)^2$, a zero point displacement of $S_{xx}^\t{zp}=(86\,\t{fm}/\rtHz)^2$, and a ground state cooling requirement $S_{xx}^\t{imp,gs}=S_{xx}^\t{zp}/n_\t{th}\approx (0.68\times 10^{-17}\,\t{m}/\rtHz)^2$.  

\begin{figure}[h!]
\centering
\includegraphics[width=\linewidth]{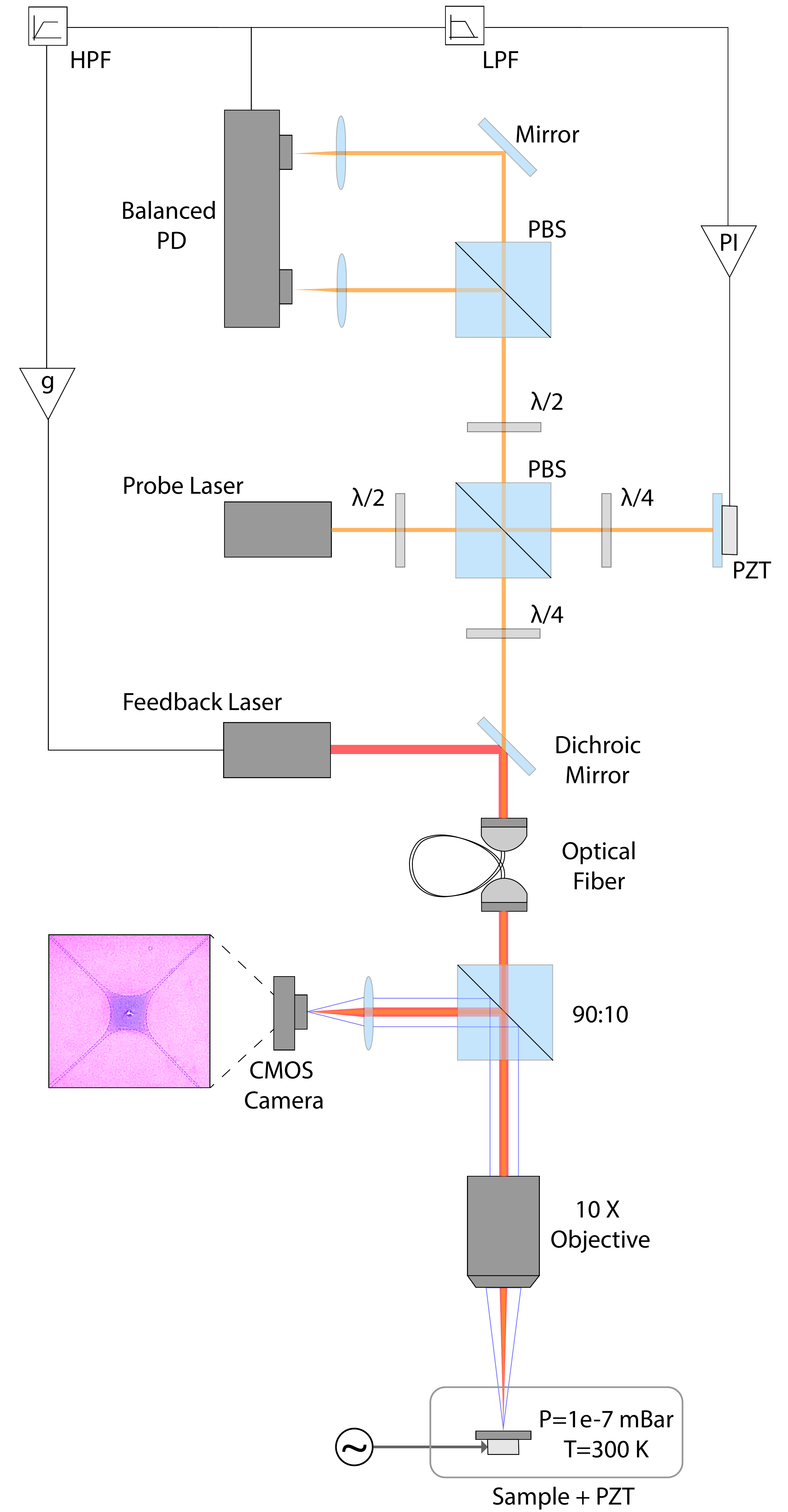}
\caption{Setup for probing the trampoline, consisting of a confocal microscope embedded in a balanced Michelsen interferometer.  Electronics for stabilizing the interferometer path length (PI = proportional integral controller, Newport LB1005) and for radiation pressure feedback cooling (see main text for details) are indicated in black. An image of the focused optical beam on the trampoline pad is shown at bottom left. }
\label{fig:interferometer}
\end{figure}

\section{Interferometric readout}

\begin{figure}[t]
\centering
\includegraphics[width=\linewidth]{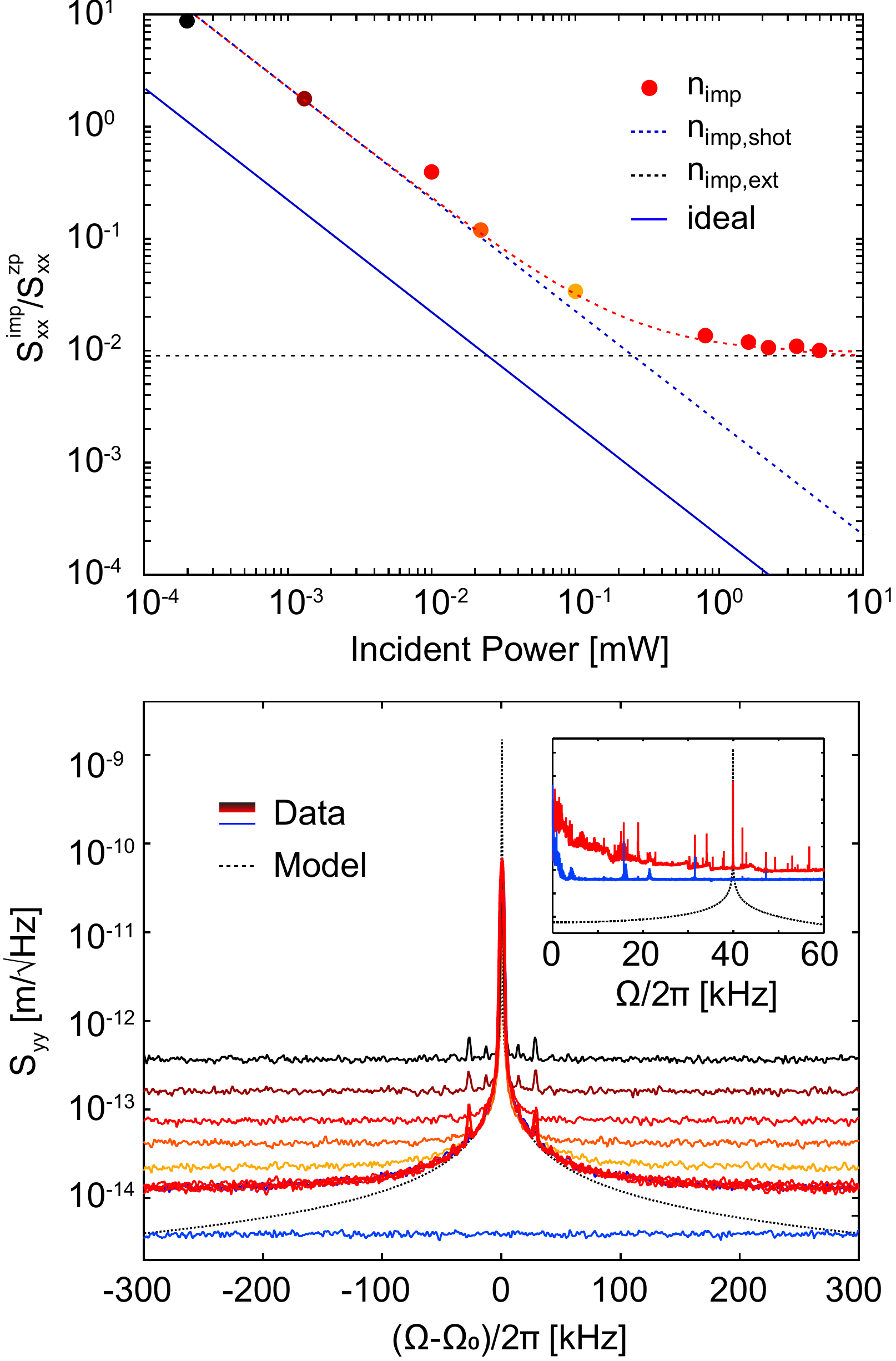}
\caption{Characterization of interferometer sensitivity.  Upper plot: Imprecision in noise quanta units versus power, compared to Eq. 8 (blue line).  Lower plot: Apparent displacement spectrum of the trampoline versus frequency for different optical powers. Dashed line is a model for $S_{xx}$. Blue line is obtained by blocking the signal arm of the interferometer at highest power.  (Inset: broadband spectrum for highest power).  }
\label{fig:feedback}
\end{figure}

Displacement of the trampoline was read out using a confocal microscope integrated into a Michelson interferometer \cite{barg2017measuring}.  Details are shown in Fig. 2. To  minimize gas damping, the device chip is mounted in a high vacuum chamber operating at $<10^{-7}$ mbar. This is enabled by a long-working-distance microscope objective (Mitutoyo M Plan APO 10X) with a spot diameter of $<5\,\mu$m.  The light source used for the experiment was an 850 nm external cavity diode laser (Newport TLB-6716).  To mitigate laser frequency and intensity noise, both the arm length and power of the interferometer were carefully balanced.

Ideally, the interferometer sensitivity is limited by shot noise
\begin{equation}
    S_{xx}^\t{imp,shot} =\frac{\hbar c\lambda}{16\pi\eta}\frac{R_\t{m}}{P}
\end{equation}
where $P$ is the power incident on the membrane, $\lambda$ is the laser wavelength, $R_\t{m}$ is the membrane reflectance, and $\eta\in [0,1]$ is the detection efficiency.  We investigated this limit by recording apparent displacement spectra $S_{yy}$ at different optical powers, where $y$ is proportional to the voltage signal produced by the balanced photoreceiver (Newport 1807).  Results are shown in Fig. 3.  To calibrate each measurement, a piezo underneath the device chip was used to drive the trampoline near its fundamental resonance with a coherent amplitude of $x_\t{cal} = 0.8$ pm (inferred by bootstrapping to the area beneath the thermal noise peak,  $\langle x^2\rangle \approx 2x_\t{zp}^2 n_\t{th}$, after an averaging time $>Q_0/\Omega_0$). At low powers ($P<100\,\mu\t{W}$), 
$S_{xx}^\t{imp}$ scales inversely with power, as expected for shot noise, with an apparent efficiency of $\eta \approx 10\%$ (using $R_\t{m} = 0.3$ \cite{wilson2009cavity}).  This value is consistent with the return loss of our microscope objective (which employs a free-space-to-fiber coupler), and in principle allows cooling to $\langle n\rangle \approx 1.1$.

In practice, the interferometer is limited by extraneous noise at sufficiently high power.  This is seen in Fig. 3 for powers above 1 mW, where the noise floor saturates to $S_{xx}^\t{imp,ext}\approx 10\, \t{fm}/\rtHz$.  Broadband measurements (Fig. 3, inset) suggest that $S_{xx}^\t{imp,ext}$ is related to differential polarization or path-length fluctuations, possibly exascerbated by peaking of the homodyne phase lock.  (We note that extraneous laser frequency noise was ruled out by introducing a path length imbalance of several millimeters, to no apparent effect.)   Although an impressive two orders of magnitude below the \emph{intrinsic} zero-point motion $(n_\t{imp,ext} \approx 0.01)$, this extraneous noise practically limits feedback cooling of the fundamental trampoline mode to $\langle n \rangle \approx 2.5\times 10^{3}$ starting at room temperature ($n_\t{th} = 1.6\times 10^8$), according to \eqref{eq:7}.  

Heating from optical absorption is another important consideration at high powers.  To investigate this effect, we consider the off-resonant thermal noise in Fig. 3, which in the presense of heating should increase linearly with power ($S_{xx}[\Omega-\Omega_0\gg\ g\Gamma_0]\propto n_\t{th}\Gamma_0/(\Omega-\Omega_0)^2$).  The variation observed is within the $\sim10\%$ statistical error of the spectral density estimate, suggesting that heating is less than $10\,\t{K}/\t{mW}$.  This value is consistent with a simple heat conduction model $dT/dP\approx \alpha L/{4wh\kappa}$, assuming a thermal conductivity of $\kappa=3\,\t{W}/\t{m\,K}$ and a conservative optical absorption coefficient of $\alpha = 10\,\t{ppm}$ \cite{wilson2009cavity}.


\begin{figure}[t]
\centering
\includegraphics[width=\linewidth]{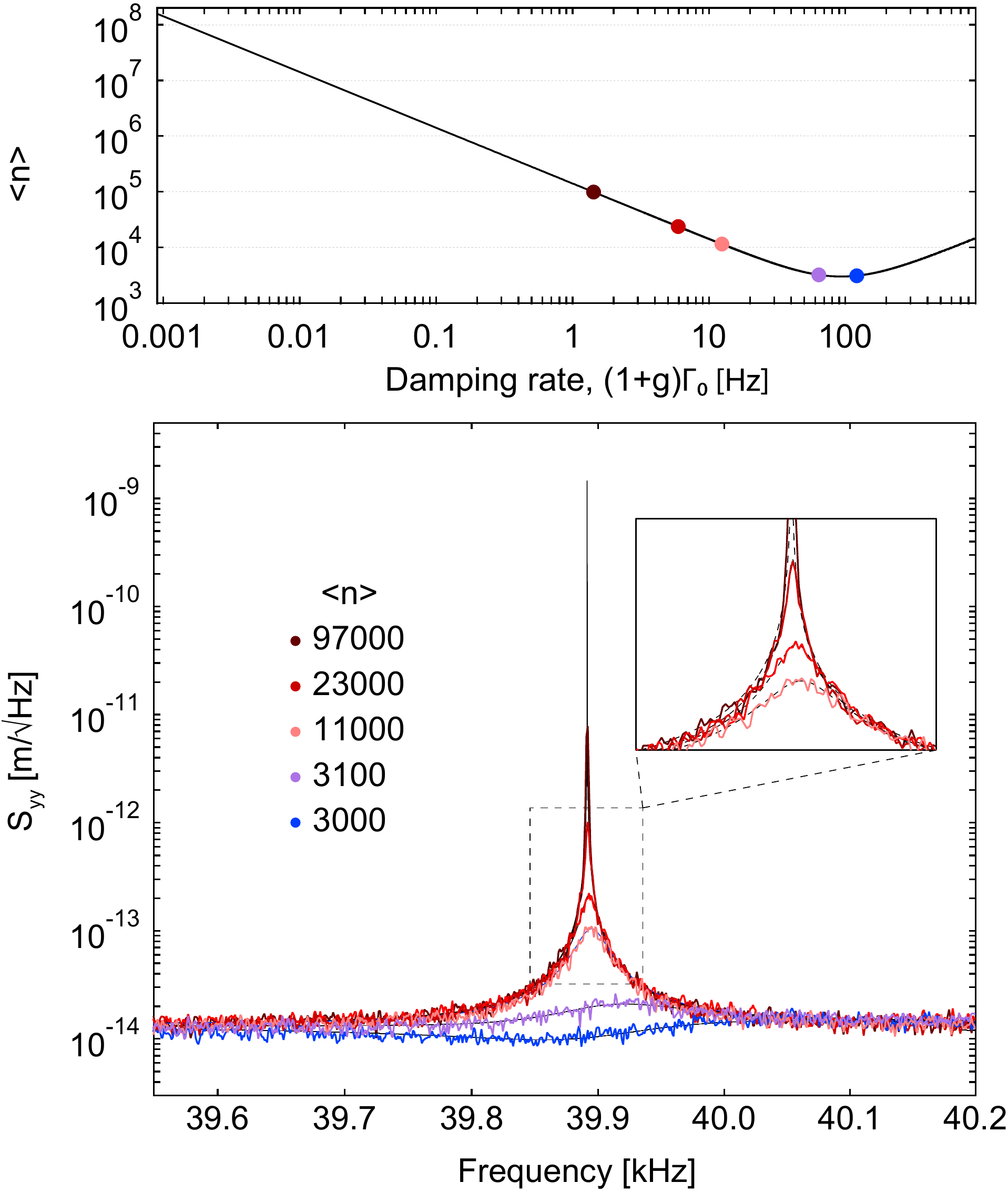}
\caption{Radiation pressure feedback cooling. Upper plot: Feedback cooling curve for parameters described in the main text.  Colored points correspond to models overlaying experimental data. Lower plot: Experimental measurements (colored) overlaid with models (dashed curves) using Eq. 9.  The solid black curve is a model for $g=0$ (no feedback).}
\label{fig:feedback}
\end{figure}

\section{Radiation Pressure Feedback cooling}

Feedback cooling was carried out using radiation pressure actuation.  The main advantage of this approach is its high bandwidth; however, we note that other methods such as piezo-electric \cite{sridaran2011electrostatic} and dielectric \cite{unterreithmeier2009universal} actuation are in principle equally viable and may be simpler to interface with feedback electronics.

To implement radiation pressure feedback, we introduce a second laser beam into the microscope which is intensity modulated by an amplified copy of the photosignal.  Specifically, we use a 670 nm laser diode (Hitachi HL6712) modulated by dithering its drive current about the threshold value.  To approximate derivate feedback while suppressing feedback to higher order modes, the photosignal is passed hrough a $10-50$ kHz bandpass filter and a delay line, resulting in an approximately $\phi = 90^\circ$ phase shift for frequencies near mechanical resonance.  The feedback force can in this case be approximated as
\begin{equation}\label{eq:9}
    \delta F_\t{fb} \approx -gm\Gamma_0 (\dot{y} + \Omega_0\cot(\phi) y) 
\end{equation}
corresponding to a normalized susceptbility $\chi_g[\Omega]^{-1}\approx (1+g)+2i(\Omega-\Omega_0)/\Gamma_0+ i g\cot(\phi)$, where $ig\cot(\phi)$ is a residual feedback stiffening term that contributes negligibly to cooling.

The results of feedback cooling with a 3 mW read out beam and a 60 $\mu$W feedback beam are shown in Fig. 4.  The feedback gain is tuned electronically using a voltage pre-amplifier (Stanford Research Systems SR560). 
To estimate $\langle n \rangle$, thermal noise spectra are fit to \eqref{eq:6} with $g$ as a free parameter, assuming $n_\t{th} = 1.56\times 10^8$, $n_\t{imp} = 0.013$, $\Gamma_0=2\pi\cdot1.5\,\t{mHz}$, and $\phi = -0.15$ (\eqref{eq:9}). To faciliate fitting, in Fig. 4, we focus on high gain settings for which the loaded damping rate $(1+g)\Gamma_0>1$ Hz. The model accurately reproduces the noise spectra until the damped peak coincides with the noise floor, for which the inferred gain is $g = 1.4\times 10^5$, corresponding to $\langle n\rangle= 3.0\times 10^{3}$.  At higher gain, the noise floor exhibits typical ``squashing'' behavior \cite{poggio2007feedback,wilson2015measurement} and the inferred $\langle n\rangle$ begins to increase.



\section{Summary and Outlook}
We have demonstrated measurement-based feedback cooling of a 40 kHz \SiN trampoline resonator from room temperature ($1.6\times10^8$ phonons) to an effective temperature of 5 mK ($3\times10^3$ phonons) using a simple two-path interferometer. 
The main limitation of our experiment is technical noise at the level of $10\,\t{fm}/\rtHz$.  Absent this noise, the apparent $10\%$ efficiency of our interferometer would in principle enable cooling to $\langle n\rangle\sim 100$ with a probe power of several mW.  Operating at 4 K, assuming no increase in mechanical Q and no photothermal heating, would enable cooling to $\langle n\rangle \sim 10$, for which motional sideband asymmetry could be readily measured.  

We speculate that a combination of monolithic interferometer design and moderate cryogenics could give access to $\langle n \rangle\sim 1$ without an optical cavity for state-of-the-art \SiN thin film resonators.  Particularly compelling are ``soft-clamped'' nanobeams \cite{ghadimi2018elastic}, which have demonstrated Megahertz modes with quality factors approaching $10^9$ and zero-point spectral densities exceeding $1\,\t{pm}/\rtHz$, and can be read out with high efficiency by evanescent coupling to optical waveguide.  For soft-clamped resonators, an important challenge is the large density of states and low thermal conductance of the phononic crystal shield, which reduces power handling capacity and can introduces extraneous thermal noise.  Clamp-optimized trampolines \cite{sadeghi2019influence,reinhardt2016ultralow, norte2016mechanical} might offer a simpler route, since the fundamental mode is well-isolated and can also have $Q_0>10^8$ at millimeter dimensions \cite{ghadimi2018elastic}.  In the future, resonators made of strained crytalline thin films promise $Q_0 >10^9$ and increased thermal conductivity at 4 K \cite{romero2019engineering}.  A recent proposal for soft-clamping fundamental modes using a ``fractal clamp" might push this performance even further \cite{fedorov2020fractal}.  

\end{document}